\documentclass[twocolumn,english,aps,pra,showpacs,preprintnumbers,floatfix]{revtex4}
\usepackage{fontenc}
\usepackage[latin9]{inputenc}
\usepackage{amsmath}
\usepackage{graphicx}
\usepackage{amssymb}
\usepackage{esint}
\usepackage{epsfig}
\usepackage{babel}

\begin{document}

\title{Two-photon ladder climbing and transition to autoresonance in a
chirped oscillator}
\author{I. Barth and L. Friedland}
\affiliation{Racah Institute of Physics, Hebrew University of Jerusalem, Jerusalem 91904,
Israel }

\begin{abstract}
The two-photon ladder climbing (successive two-photon Landau-Zener-type
transitions) in a chirped quantum nonlinear oscillator and its classical
limit (subharmonic autoresonance) are discussed. An isomorphism between the
chirped quantum-mechanical one and two-photon resonances in the system is
used in calculating the threshold for the phase-locking
transition in both the classical and quantum limits. The theory is tested by
solving the Schrodinger equation in the energy basis and illustrated via the
Wigner function in phase space.
\end{abstract}

\date{\today }
\maketitle

The transition between the quantum and classical descriptions of dynamical
systems played a pivotal role in the foundation of quantum mechanics. In
this context, the correspondence principle addressed the classical limit of
a quantum system for large quantum numbers \cite{Bohr}, such that the
classical equations of motion describe the average wave packet \cite%
{Ehrenfest}. Since these early works, studying subtitles of the
quantum-classical crossover still comprises a field of active research (e.g.
\cite{Littlejohn,Gutzwiller,Kippenberg,Maeda}). An instructive framework for
theoretical and experimental investigation of this correspondence is the
ac-driven non-linear oscillator. Recent studies in the field involved
nonlinear resonators in a nanoelectromechanical system \cite{Katz},
parametrically modulated oscillators \cite{Dykman}, and chirped-driven
Josephson junctions \cite{ido4,ido6Yoni}. Here we focus on the
quantum-classical transition in an oscillator exhibiting the classical
\textit{subharmonic\ autoresonance} phenomenon, i.e. a continuing
phase-locking with a driving perturbation slowly passing, say, 1/2 the
natural frequency of the oscillator.

Autoresonance (AR) is a continuing phase-locking between a classical
nonlinear oscillatory system and a chirped frequency driving perturbation.
The phenomenon was first utilized in relativistic particle accelerators \cite%
{Livingstone}. In the last two decades, AR was recognized as a robust method
of excitation and control of nonlinear systems, ranging from atoms \cite%
{lazar62} and molecules \cite{Liu} through plasmas \cite{Danielson,lazar131}
and fluids \cite{Ben-David}, to nonlinear optics \cite{lazar129}. The most
recent applications involved anti-hydrogen project at CERN \cite{ALPHA PRL}
and superconducting Josephson junctions \cite{lazar121,ido4,ido6Yoni}. The
salient feature of the AR is a sharp threshold for capture into resonance by
passage through the fundamental linear resonance \cite{lazar85}. The width
of this threshold depends on the temperature of the initial state \cite{ido3}%
, while in the low temperature limit, this width saturates to a finite value
associated with the zero-point fluctuations of the quantum ground state \cite%
{ido4,ido5}.

The quantum counterpart of the AR is the ladder climbing (LC), characterized
by continuing successive two-level Landau-Zener \cite{LZ} transitions. This
process was studied by Marcus et al. \cite{Giladtheory,Giladexp} in
application to driven molecules, where chirped frequency laser radiation
resonantly interacts with successive energy gaps of the molecule. In
addition, the LC was studied in the context of Morse oscillator \cite{Guerin}
and more recently in Josephson junctions \cite{ido6Yoni} and Rydberg atoms
\cite{Jiang}. The transition between the classical AR and the quantum LC was
studied in \cite{ido5,Giladtheory}.

The classical subharmonic autoresonance (SHAR) is the phase-locked response
of a nonlinear oscillator to a chirped driving force passing through a
rational fraction of the fundamental linear frequency. This phenomenon was
studied in classical nonlinear oscillators \cite{lazar88} and plasmas \cite%
{lazar89}. On the other hand, quantum multiphoton processes were studied
both experimentally and theoretically via adiabatic Floquet analysis in
association with atomic systems \cite{Baruch,Gatzke,Conover,Chang}, but the
issue of the quantum counterpart of the classical SHAR in a driven chirped
nonlinear oscillator was not addressed previously. These processes may be
important in such applications as quantum Josephson circuits and
nanomechanical systems.

Here we discuss this problem for the first time and show that the quantum
counterpart of this process is indeed the multiphoton ladder climbing
(MPLC). We will use the isomorphism between the fundamental and the
subharmonic $1:2$ autoresonances (the generalization to the $1:n$ resonances
can be obtained similarly) to estimate the chirped SH resonant capture
probability in both the classical and quantum limits and compare our
predictions with numerical simulations.

We focus on a driven weakly nonlinear oscillator governed by the
dimensionless Hamiltonian
\begin{equation}
H=\frac{1}{2}\left( p^{2}+x^{2}\right) +\frac{1}{3}\lambda x^{3}+\frac{1}{4}%
\beta x^{4}+\varepsilon x\cos \varphi _{d},  \label{eq:Hamiltonian}
\end{equation}%
where $\varphi _{d}$ is the driving phase, such that the driving frequency $%
\omega _{d}(t)=d\varphi _{d}/dt$ is a slowly varying function of time. The
classical \textit{fundamental} AR and the corresponding quantum LC processes
in the problem are associated with the case, when the driving frequency
passes through the fundamental linear frequency of the oscillator, e.g. $%
\omega _{d}(t)=1+\alpha t,$ $\alpha $ being the chirp rate. This problem was
studied quantum mechanically in Refs. \cite{Giladtheory,ido5}. The analysis
was based on the expansion of the wave function of the oscillator, $|\psi
\rangle =\sum_{n}c_{n}|\psi _{n}\rangle $, in the energy basis $|\psi
_{n}\rangle $ of the undriven Hamiltonian i.e., $H(\varepsilon =0)|\psi
_{n}\rangle =E_{n}|\psi _{n}\rangle ,$ where $\langle \psi _{k}|\psi
_{n}\rangle =\delta _{k,n}$. In this basis, the dimensionless ($\hbar =1$)
Schrodinger equation yields
\begin{equation}
i\frac{dc_{n}}{dt}=E_{n}c_{n}+\varepsilon \sum_{k}c_{k}\langle \psi _{k}|%
\hat{x}|\psi _{n}\rangle \cos \varphi _{d},  \label{eq:schrodinger}
\end{equation}%
The energy levels in (\ref{eq:schrodinger}) for sufficiently small $n$ can
be approximated as \cite{Landau QM}
\begin{equation}
E_{n}\approx n+\frac{1}{2}+\gamma (n^{2}+n)+\frac{3}{16}\beta -\frac{11}{72}%
\lambda ^{2},  \label{En}
\end{equation}%
$n=0,1,2,...$, and $\gamma =\frac{3}{8}\beta -\frac{5}{12}\lambda ^{2}$. The
linear approximation
\begin{equation}
\langle \psi _{k}|\hat{x}|\psi _{n}\rangle \approx \frac{\sqrt{n}\delta
_{k,n-1}+\sqrt{n+1}\delta _{k,n+1}}{\sqrt{2}}\equiv K_{kn}^{L}
\label{KLinear}
\end{equation}%
for the coupling terms in Eq. (\ref{eq:schrodinger}) was used in Ref. \cite%
{ido5} in analyzing the passage though the fundamental resonance in the
problem. One can define three characteristic times in the problem of passage
through the fundamental resonance \cite{Giladtheory,ido5}, i.e. $%
T_{NL}=2\gamma /\alpha $ (the time of passage through the nonlinear
frequency shift between the first two transitions on the energy ladder), $%
T_{R}=\sqrt{2}/\varepsilon $ (the inverse Rabi frequency), and $T_{S}=1/%
\sqrt{\alpha }$ (the frequency sweep time scale). These three times yielded
two dimensionless parameters: $P_{1}=T_{S}/T_{R}=\varepsilon /\sqrt{2\alpha }
$ (measuring the strength of the drive) and $P_{2}=T_{NL}/T_{S}=2\gamma /%
\sqrt{\alpha }$ (characterizing the nonlinearity). It was shown that $P_{1,2}
$ fully characterize the phase-locking transition in the fundamental
resonance case \cite{ido5}. In order to address the problem of two-photon LC
(slow passage through $1:2$ resonance), we use
\begin{equation}
\omega _{d}(t)=\frac{1}{2}\left( 1+\alpha t\right) .  \label{driving frq}
\end{equation}%
Numerical solutions of Eq. (\ref{eq:schrodinger}) in this case with the
coupling terms of Eq. (\ref{KLinear}) show no two-photon transition. This
different process requires inclusion of additional higher order coupling
terms associated with the nonlinearity. Consequently, we replace Eq. (\ref%
{KLinear}) by
\begin{equation}
\langle \psi _{k}|\hat{x}|\psi _{n}\rangle \approx K_{kn}^{L}+\lambda
Q_{kn}+\beta R_{kn},  \label{Full}
\end{equation}%
where, by standard perturbation theory \cite{Landau QM},
\begin{eqnarray*}
Q_{kn} &=&\frac{1}{6}[-3\left( 2n+1\right) \delta _{k,n}+\sqrt{(n+1)(n+2)}%
\delta _{k,n+2} \\
&&+\sqrt{n(n-1)}\delta _{k,n-2}]
\end{eqnarray*}%
and
\begin{eqnarray*}
R_{kn} &=&\frac{1}{24\sqrt{2}}[3\sqrt{(n+1)(n+2)(n+3)}\delta _{k,n+3} \\
&&-2\left( 2n+3\right) \sqrt{(n+1)}\delta _{k,n+1}-2\left( 2n+1\right) \sqrt{%
n}\delta _{k,n-1} \\
&&+3\sqrt{n\left( n-1\right) \left( n-2\right) }\delta _{k,n-3}].
\end{eqnarray*}

\begin{figure}[tp]
\includegraphics[width=9cm]{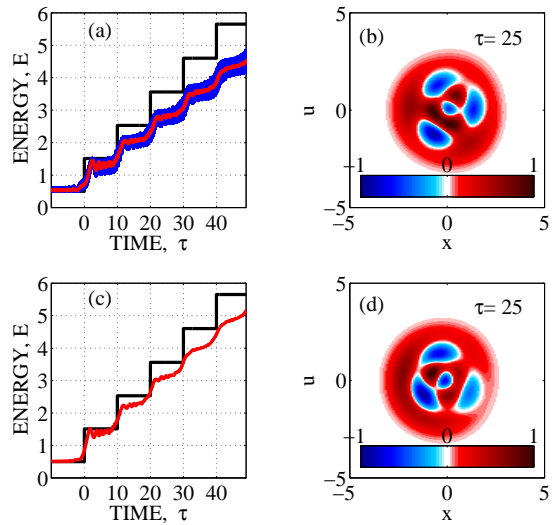}
\caption{(color online) The dynamics in the energy basis (a,c) and the
corresponding Wigner functions (b,d) at time $\protect\tau =25$ in the $1:2$
subharmonic (a,b) and the fundamental (c,d) quantum ladder climbing regime,
with the same $P_{2}=10$, but $P_{1}$ divided by $\protect\epsilon \protect%
\lambda $.}
\label{Flo:fig1}
\end{figure}

At this stage, we illustrate the SHLC and SHAR in simulations. We have
solved Eq. (\ref{eq:schrodinger}) numerically, subject to ground state
initial conditions, $c_{n}\left( t_{0}=-10/\sqrt{\alpha }\right) =\delta
_{n,0}$, for two sets of parameters, in the quantum SHLC (Fig.1) and the
classical SHAR (Fig. 2) regimes. Figure 1a corresponds to the set of
parameters $\{\alpha ,\beta ,\lambda ,\varepsilon
\}=\{10^{-6},0.016,0.05,0.18\}$ and shows the energy of the system versus
the slow time $\tau =\sqrt{\alpha }t$. Taking $40$ levels into account \ was
sufficient in this example. One can see that the response of the quantum
nonlinear oscillator to the chirped frequency drive is by successive
transitions between neighboring energy levels. The red line in the figure is
the time average over an interval of $\Delta \tau =0.1,$ eliminating fast
oscillations in the dynamics, similar to the procedure used in ref. \cite%
{lazar88}. The theoretical, perfect energy ladder climbing scenario is
illustrated in Fig. 1a by the solid black line. We also observe that,
similar to the fundamental ladder climbing, the nonlinearity parameter\ $%
P_{2}=2\gamma /\sqrt{\alpha }=10$ in the SHLC regime is much larger than
unity and that the transitions between neighboring levels occur at times, $%
\tau _{n}=nP_{2}$ \cite{ido5}. For further illustration, we have calculated
the Wigner function \cite{Schleich} in phase space and show a snapshot of
time $\tau =25$ in Fig. 1b. The Wigner function exhibits structure
characteristic to the $n=3$ level of the quantum ladder as is expected at
this time from Fig. 1a, while the probability of capture into resonance
(total occupation of resonant levels) was $74\%$. It is instructive to
compare these results with those for the fundamental LC case presented in
Figs. 1c,d and obtained by using the same set of parameters, but $%
\varepsilon $ replaced by $\varepsilon ^{2}\lambda $ ($P_{1}$ multiplied by $%
\varepsilon \lambda $). The resonant capture probability in\ this case was $%
84\%$, while Fig. 1 exhibits a similarity between the simulations results
for the fundamental and SH resonances with this choice of parameters.

The second numerical example shown in Fig. 2a,b uses the same initial
conditions, but parameters $\{\alpha ,\beta ,\lambda ,\varepsilon
\}=\{10^{-4},0.0016,0.0155,1.9\}$, and the calculation involves $250$
quantum levels. Here, $P_{2}=0.1$ describing the classical limit ($P_{2}\ll
1 $) \cite{ido5}, where the energy does not vary in steps, but grows
monotonically with superimposed slow oscillations, as expected from the
theory of the classical nonlinear resonance \cite{Sagdeev}. As above, the
thin red line in the figure is the time average of the results over a window
of $\Delta \tau =0.1$ for eliminating fast oscillations. A snapshot of the
calculated Wigner function in this example at time $\tau =6$ is shown in Fig
2b and the probability of capture into AR was $85\%$. The figure shows that
the most populated part of the phase space is a crescent corresponding to
the resonantly trapped phase space area of the oscillator, while the
characteristic interference patterns (which can be eliminated by coarse
graining) is seen in nonresonant regions of phase space. As in the previous
LC example, we compare these results with the corresponding fundamental
resonance case shown in Figs. 2c (energy evolution) and 2d (Wigner
function), where $P_{2}$ is the same, but $P_{1}$ again multiplied by the
factor $\varepsilon \lambda $. The capture probability in this case was $%
99\% $. One observes again a noticeable similarity between the SH and
rescaled fundamental autoresonance cases.

\begin{figure}[tp]
\includegraphics[width=9cm]{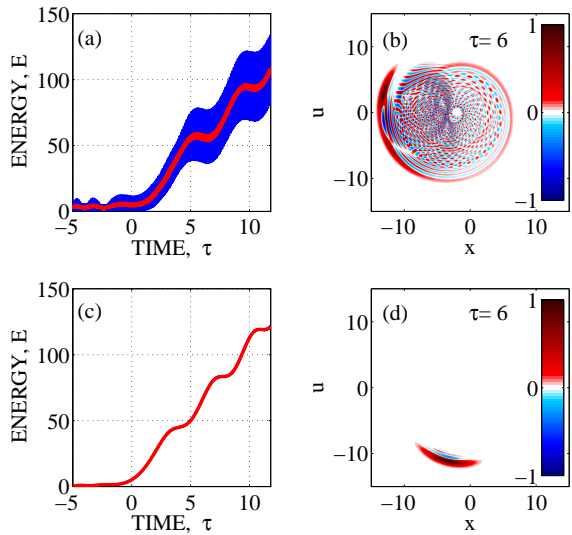}
\caption{(color online) The dynamics in the energy basis (a,c) and the
corresponding Wigner function (b,d) at time $\protect\tau =6$ in the $1:2$
subharmonic (a,b) and the fundamental (c,d) classical autoresonance regime,
with the same $P_{2}=0.1$, but $P_{1}$ divided by $\protect\epsilon\protect%
\lambda$.}
\label{Flo:fig2}
\end{figure}

Our theoretical analysis uses the following canonical transformation of the
coordinate and momentum
\begin{eqnarray}
x^{\prime } &=&e^{-iS}xe^{iS}  \notag \\
p^{\prime } &=&e^{-iS}pe^{iS},
\end{eqnarray}%
where $S=\frac{4}{3}\varepsilon \left( x\sin \varphi _{d}+p\cos \varphi
_{d}\right) $. The transformed Hamiltonian in this case becomes
\begin{equation}
H^{\prime }=e^{-iS}He^{iS}+\frac{dS}{dt}
\end{equation}%
where, as before, we set $(\hbar =1)$. The first term in the RHS can be
calculated via the identity \cite{Landau QM}
\begin{equation}
e^{-iS}He^{iS}=H-i[S,H]-\frac{1}{2}[S,[S,H]]+...
\end{equation}%
Then, one finds that all $O(\varepsilon )$ terms in the transformed
Hamiltonian, $H^{\prime }$, vanish. We seek $1:2$ subharmonic resonance in
the problem as the driving frequency $\omega _{d}=\dot{\varphi _{d}}\approx
\frac{1}{2}$ passes through the two photon resonance. There exist only one $%
O(\varepsilon ^{2})$ two photon resonant term in the transformed
Hamiltonian, i.e. $\frac{8}{9}\varepsilon ^{2}\lambda x^{\prime } \cos 2\varphi _{d}$.
After neglecting all other nonresonant and higher order terms, the
transformed Hamiltonian becomes
\begin{equation}
H^{\prime }=\frac{1}{2}\left( p^{\prime 2}+x^{\prime 2}\right) +\frac{1}{3}%
\lambda x^{\prime 3}+\frac{1}{4}\beta x^{\prime 4}+\frac{8}{9}\varepsilon
^{2}\lambda x^{\prime }\cos 2\varphi _{d}.
\label{eq:transformed_Hamiltonian}
\end{equation}%
One can see that this Hamiltonian with $\omega _{d}=\dot{\varphi _{d}}=\frac{%
1}{2}\left( 1+\alpha t\right) $ is the same as the Hamiltonian (\ref%
{eq:Hamiltonian}) with $\omega _{d}=\dot{\varphi _{d}}=1+\alpha t$ studied
in ref. \cite{ido5} for the fundamental resonance case, but with $%
\varepsilon $ replaced by $\frac{8}{9}\varepsilon ^{2}\lambda $. This
explains the similarity between the chirped fundamental and the SH
autoresonance, illustrated in our numerical examples. The same isomorphism
was found in the classical theory of the SHAR \cite{lazar88}. Consequently,
the parameter $P_{1}$ for the fundumental resobnance should be replaced by $%
\widetilde{P}_{1}=\frac{8}{9}\varepsilon \lambda P_{1}$ for the two-photon
resonance case, while $P_{2}$ remains unchanged. Note that the classical
problem of the fundamental AR is fully controlled by a \textit{single}
parameter, $\mu =\frac{1}{2}P_{1}P_{2}^{1/2}$ \cite{ido5} ($\widetilde{\mu }=%
\frac{1}{2}\widetilde{P}_{1}P_{2}^{1/2}$ in the subharmonic case). In
contrast, the quantum mechanical counterpart in the problem is characterized
by two parameters ($\widetilde{P}_{1}$,$P_{2}$) due to a new scale
associated with $\hbar $.

\begin{figure}[tp]
\includegraphics[width=7cm]{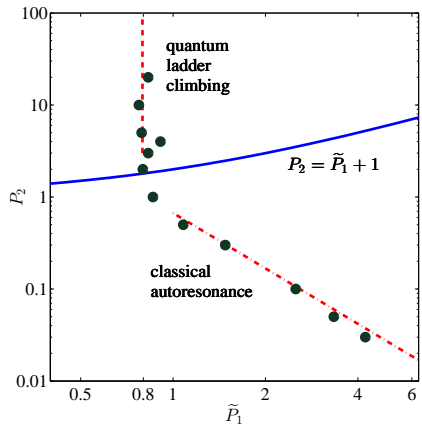}
\caption{(color online) Different regimes of the phase-locking transition in
the chirped, $1:2$ subharmonic resonance. The dots show the location of the
threshold for the phase-locking transition. The dashed and dashed-dotted
lines represent the theoretical thresholds in the quantum SHLC and classical
SHAR regimes, respectively, while the solid line separates these regimes.}
\label{Flo:fig3}
\end{figure}%

The aforementioned isomorphism allows to apply all the results of the theory
of the fundamental chirped resonance to the SH scenario by replacing $%
P_{1}\rightarrow \widetilde{P}_{1}$. For instance, in the fundamental
resonance case, it was found that the separator between the classical and
the quantum regimes in the $(P_{1},P_{2})$ parameter space is the line $%
P_{2}=P_{1}+1$ \cite{ido5}. Hence, we conclude that in the SH case, the
classicality condition is $P_{2}<\widetilde{P}_{1}+1$. Furthermore, the
probability of capture into fundamental resonance depends on the parameters $%
P_{1,2}$ and, thus, in the SH case, this probability is fully described by
the parameters $\widetilde{P}_{1},P_{2}$. For example, the threshold for the
phase-locking transition by passage through the fundamental resonance is a
line in the parameter space $P_{1cr}=f(P_{2})$ defined as the value of $%
P_{1} $ for which the capture probability is $50\%$ \cite{ido5}. Therefore,
in the SH case the equivalent threshold line is $\widetilde{P}%
_{1cr}=f(P_{2}) $. Figure \ref{Flo:fig3} compares these predictions with the
results of the numerical solution of Eq. (\ref{eq:schrodinger}) for
different values of $P_{2}$. The dots in the figure show the numerically
found threshold, while the dashed lines correspond to the appropriately
rescaled theoretical predictions in the classical SHAR, i.e. $\widetilde{P}%
_{1cr}=0.82/\sqrt{P_{2}}$ (dashed doted line), and in the quantum SHLC, $%
\widetilde{P}_{1cr}=0.79$ (dashed line) limits rescaled from the fundamental
resonance theory \cite{ido5}. One observes a very good agreement in both
dynamical limits, including the characteristic transition between these two
limits near the rescaled theoretical separator, $P_{2}=\widetilde{P}_{1}+1$
(solid line).

In conclusion, we have studied the problem of passage through two-photon
nonlinear resonance and identified the quantum counterpart of the classical
SHAR in the nonlinear oscillator, i.e. the quantum two-photon ladder
climbing. We have used the isomorphism between the quantum fundamental and
the two-photon chirped resonance phenomena. A similar isomorphism for
stationary $(\alpha =0)$ quantum resonance exists as a special case of the
chirped resonance. The calculation can be generalized to similar $n>2$
photon processes. The theory shows that all the results in the chirped
fundamental resonance process in both quantum and classical limits can be
extended to the subharmonic resonance case by simply rescaling the driving
parameter. The engineering and control of a desired quantum state of the
oscillator via the ladder climbing process can be achieved by passage
through both the fundamental and SH resonances. However, in the $1:2$
subharmonic chirped resonance case, a desired state is achieved with just $%
1/2$ of the driving frequency bandwidth, as compared to the same final state
reached via passage through the fundamental resonance.

The authors are grateful to O. Gat for useful discussions. Supported by the
Israel Science Foundation (Grant No. 451/10).

\end{document}